\documentclass[twocolumn,showpacs,preprintnumbers,amsmath,amssymb]{revtex4}
\usepackage{dcolumn}
\usepackage{color}
\usepackage{bm}
\usepackage{graphicx}
\def \be{\begin{equation}}
\def \ee{\end{equation}}
\def \bea{\begin{eqnarray}}
\def \eea{\end{eqnarray}}

\def \tq{{{\tilde q}}}
\def \cH{{\cal H}}
\def \cA{{\cal A}}
\def \cV{{\cal V}}
\def \cO{{\cal O}}

\def \txi{{\tilde \xi}}

\def\bE{{\bf E}}

\def\bR{{\bf R}}
\def\ba{{\bf a}}
\def\bp{{\bf p}}
\def\bq{{\bf q}}
\def\hbq{{\hat{{\bf q}}}}
\def\bx{{\bf x}}
\def\by{{\bf y}}
\def\bz{{\bf z}}

\def\br{{\bf r}}
\def\bj{{\bf j}}
\def\half{{1\over 2}}
\def\etal{{\em et  al.}}
\def \bPi{\mbox{\boldmath{$\pi$}}}
\def \bxi{\mbox{\boldmath{$\xi$}}}
\def\etal{{\em et.~al.}}
\begin{document}
\title{ Nonlinear Current of Strongly Irradiated Quantum Hall Gas}
\author{Assa Auerbach and G. Venketeswara Pai}
\address{Physics Department, Technion, Haifa  32000, Israel}
\begin{abstract}
Two dimensional electrons  in weakly disordered high Landau levels are considered.
The current-field response in the presence of a strong microwave field, is computed.
The {\em disordered  Floquet evolution operator} allows us  to treat the short range disorder  perturbatively, at any strength of  electric fields.  
A simplifying {\em Random Matrix Approximation} reproduces the broadened Landau levels density of states and structure factor. 
We derive the magnitude of the Microwave Induced Resistivity Oscillations. The disorder short wavelength cut-off 
determines the non-linear electric fields of the Zero Resistance State  and the Hall Induced  Resistivity Oscillations.  
We discuss  wider implications of our results on experiments and other theories.
 \end{abstract}
\date{\today}
{\Large \bf}
\maketitle
\vskip2pc 
\narrowtext
\section{Introduction}
Some remarkable phenomena have been recently observed in high mobility GaAs/AlGaAs hetero-structures:
Microwave Induced Resistance Oscillations (MIRO),  Zero Resistance States (ZRS) \cite{Mani,Zudov1,Ye,Zudov2,Studenkin}, 
and  Hall Induced  Resistivity Oscillations (HIRO)\cite{hiro1A, hiro1B,hiro1C,hiro1D,hiro1E}.  These experiments are carried out at
weak fields (relative to the Quantum Hall regime) and at temperatures $T\ge \hbar\omega_c$, ($\omega_c$ is the cyclotron frequency),
where Shubnikov-de Haas oscillations are thermally smeared.  Nevertheless,  microwave radiation, or large Hall currents, can
expose the underlying Landau quantization via the MIRO and HIRO oscillations.

MIRO (in the Hall bar geometry) exhibits large  magneto resistance oscillations as a function of radiation frequency $\omega$, with nodes at harmonics of $m\omega_c, m=1,2,\ldots$. In strong enough radiation,  the resistance in the positive detuning regimes   may be nearly completely  suppressed. This  phenomenon is commonly denoted  ZRS.
The ZRS  has been attributed  to
spontaneous generation of internal electric field domains\cite{Andreev}. The macroscopic structure and stability of ZRS domains have been investigated by a phenomenological Lyapunov functional \cite{A1,A2}.  

Microscopic theories for MIRO and ZRS, divide into two categories: (i)   The Displacement Photocurrent (DP) mechanism,
proposed by Ryzhii \etal \cite{Ryzhii} and Durst \etal \cite{Durst}. (See also \cite{Lei,Torres,Vavilov}). The DP is stronger for well resolved Landau levels, whose width $\Gamma$ is smaller than Landau level separation $\hbar\omega_c$. (ii) The Distribution Function (DF) mechanism, proposed by Dmitriev \etal, \cite{Dmitriev}. The DF mechanism
becomes important when the inelastic lifetime is much longer than the elastic transport scattering time. A recent theory of HIRO for overlapping Landau levels was given in Ref. \cite{VAG}.

Independent of the particular dominant mechanism, the disorder potential is essential for the dissipative currents. Treating disorder in high Landau  levels is an old theoretical challenge:
Since the clean system has macroscopic Landau level degeneracies, a straightforward perturbation theory in disorder is ill posed. The {\em linear, dark conductivity},  (in the absence of radiation), has been calculated by the
Self Consistent Born Approximation (SCBA)\cite{AU,comm:SCBA}. The SCBA is a selective diagram resummation, whose neglected vertex corrections are controlled by the smoothness of the disorder\cite{RS}. 
Unfortunately, a controlled extension of the SCBA to strong static and time dependent fields, has so far proven difficult. Nevertheless,   the  ZRS and the HIRO are inherently non linear effects which require theoretical attention.

In this paper we devise a new `divide and conquer' method \cite{Dietel} which  incorporates strong electric fields with the disorder potential.
It has long been appreciated 
that strong radiation effects on the {\em clean} Landau levels, are  tractable 
using a {\em Floquet transformation} \cite{Ryzhii,Torres}. 
However,  writing down an explicit Floquet transformation for a disordered Hamiltonian was believed to be intractable.
Here we make progress by exploiting the commutation between Landau  and guiding center operators.
We construct a separable  disordered Hamiltonian, which is completely 'Floquet transformable'.
This trick allows us to eliminate  the electric fields, and obtain the random Floquet (quasi-stationary) eigenfunctions. These states, describe our zeroth order densiity matrix,
without any  dissipative current.
The dissipative current is subsequently computed perturbatively, to leading order in the
remainder short wavelength disorder. The physical small parameter is the ratio of transport scattering rate  to Landau level width. 

This approach allows us to answer some outstanding questions: 
\begin{itemize}
\item  {\em What is optimal  disorder  for large MIRO and ZRS  effects?}

We identify the 'figure of merit'  as the  ratio $R=\hbar\omega_c/\Gamma$, which is also expected to be large in systems exhibiting the HIRO
effect.

\item {\em What characteristic Hall fields determine the HIRO effect?} 

The low intra-band field  is  
\be
E_\Gamma=\Gamma/ (el_B^2q_s),
\label{EG}
\ee
 and
the higher 'inter-band'  field periodicity  is given by  
\be
E_{\omega_c}=\hbar\omega_c/ (el_B^2q_s).
\label{EO}
\ee
 $q_s$ is the high wave-vector cut-off of the disorder fluctuations, and $l_B$ is the Landau length.

\item {\em What is the   expected magnitude of the spontaneous ZRS  fields?} 

The quantity  $E_\Gamma$ sets the overall scale of the ZRS field $E^{zrs}$, which also depends on $R$, frequency and microwave power.
$E^{zrs}$ has a second order dynamical phase transition at a threshold microwave power.
\end{itemize}

This paper is organized as follows: Section \ref{sec:transport}  briefly reviews macroscopic transport theory
of the Corbino and Hall bar geometries. The main purpose is to define the {\em dissipative current}, which is the subject of subsequent sections. Section  \ref{sec:Model}  introduces
the  microscopic model: Non interacting electrons in a weak magnetic feld, with   `broad- tail'  disorder correlations.
The disorder potential is split into two operators, where the long wavelength
components are (mostly) incorporated into the broadened Landau levels Hamiltonian. In section \ref{sec:Floquet}, the disordered Floquet evolution operator $U_d(t)$ is constructed.  We show that while the Landau levels broaden into random Floquet eigenstates, the DC field  (surprisingly) produces no dissipative current and a  perfect classical Hall current. The Random Matrix Approximation (RMA)   captures the spectrum and eigenfunctions of the broadened Landau levels, and recovers  the well known Self Consistent Born Approximation (SCBA).
In section \ref{sec:current},  we derive  the leading
order dissipative non linear current, Eq.~(\ref{current-final}), which can be disorder averaged numerically.
 In section \ref{sec:Results} the current formula is simplified by the RMA to a tractable analytical expression Eq.~(\ref{jx-asympt}).
The simplified expression is analyzed in some detail:  the SCBA dark conductivity is recovered, the MIRO, ZRS and HIRO effects are obtained. Predictions are obtained for the
magnitudes of these effects and the values of non linear field scales.  
A plot of the full microwave irradiated current as a function of DC field is given in Fig.~\ref{fig:JE}.
We conclude with a brief discussion of  theoretical and experimental issues pertinent to our results.

Appendix  \ref{App:disorder} calculates the disorder matrix elements, and
Appendix  \ref{App:Floquet} derives the explicit Floquet operator.

\section{Macroscopic Transport Theory}
\label{sec:transport}
Here we briefly review magneto-transport theory of the Corbino and Hall bar geometries. 
This will provide a direct relation between the experimentally
measurable currents and voltages, and the quantity calculated theoretically in later sections: {\em the non linear dissipative current} $j^d(E)$.
  
In the presence of a microwave field $\bE_\omega$, and a DC field  $\bE =E  \hat{\bE}$,  macroscopic transport theory assumes a  {\em local}  
relation  between the DC current density and the electric field,
 \bea
\bj &=& j^{d}\left(E,\bE_\omega\right)~\hat{ \bE} ~+~\bj^H(\bE), \nonumber\\
\bj^H &=&   \sigma_{H}  \hat{\bz}\times \bE.
\label{curr}\eea
$\bj^H$ is the non dissipative Hall current,  and   $\sigma_H$ is the Hall conductivity.
$\sigma_H$ and $j^d(E)$   require  input from a microscopic theory which involve
quantum mechanical scattering  processes on length scales shorter than the inelastic dephasing length.

In the presence of large scale inhomogeneities, (considered in Refs.~\cite{A1,A2}) , the dissipative current $\bj^d(\br)$ is not necessarily parallel to $\bE(\br)$, and $\sigma_H$ may 
vary in space. Here we do not consider such large scale disorder.

\subsection{Corbino Geometry}
\begin{figure}[htb]
\begin{center}
\includegraphics[width=9cm,angle=0]{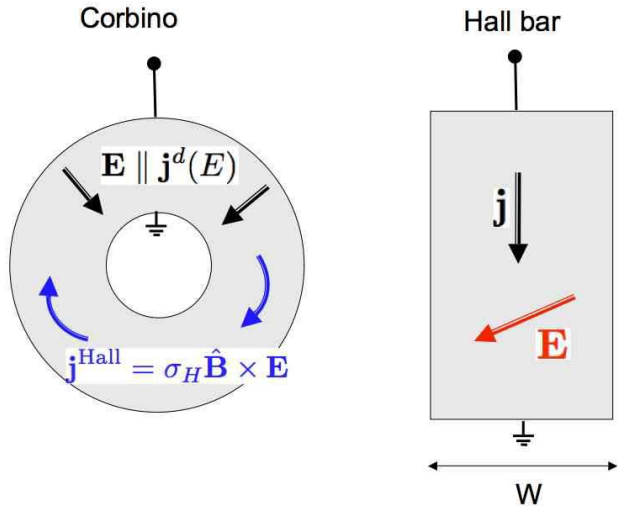}
\caption{Corbino and Hall Bar Geometries. In the homogeneous Corbino sample, the electric field and dissipative current density are radial, 
while the non dissipative Hall curent is  azimuthal. In the Hall bar, the current between leads includes both dissipative and non dissipative components. For large Hall angle $\sigma_{H}>>\sigma_{xx}$, most (but not all) of the current through the bar is non dissipative. } 
\label{fig:Geometry}
\end{center}
\end{figure}

By the definition (\ref{curr}), for a position independent   $\sigma_H$,  the dissipative and Hall currents are conserved separately:
\be
\nabla\cdot \bj= \nabla \cdot\bj^H=\nabla\cdot \left( j^d \hat{\bE}\right)= 0.
\ee
In a Corbino geometry (see Fig.\ref{fig:Geometry}), one specifies the potential on the inner and
outer boundaries of the sample which determines the purely radial electric field $\bE=E(r) \hat{\br}$. The dissipative current 
$\bj^d$ is  also radial,
while the non dissipative Hall current circulates around the annulus unrestricted. 

The Corbino geometry completely separates the effects
of $\sigma_{H}$ and $\bj^d$, and is simpler to analyze  theoretically. The function $j^d(E)$ is uniquely determined for  particular sample dimensions, from the total current and voltage between the outer  and inner edges, with no dependence on $\sigma_H$.

As was shown in Refs.\cite{Andreev,A1,A2}, in the absence of an external bias current,
a negative conductivity in some field range $j^d(E) <0$ may be unstable to a
spontaneously created  domains of finite internal  fields whose magnitudes are $|\bE|=E^{zrs}$ and orientation is determined
by the boundary conditions and long range potentials.
\be
j^d(E^{zrs})=0.
\label{ZCS}
\ee
In the absence of 'pinning'  in the Lyapunov functional, the domian walls are free to move and absorb any charge in external bias voltage
leaving the current vanishingly small. This is the so-called  'Zero Resistance State' (ZRS), which, confusingly, yields zero {\em conductance}  in the Corbino geometry.

\subsection{Hall Bar Geometry}
Experimentally, the Hall bar geometry   (see Fig.\ref{fig:Geometry}) is more popular for ease of fabrication.
The Hall bar geometry imposes $j_y=0$.
The current is forced through in one direction $\bj =j \hat{\bx}$, and the longitudinal and transverse voltage drops are measured.  
The electrochemical field  $\bE=(E_x,E_y)$ can be deduced from the geometry.
The can be deduced.  By (\ref{curr}),  
the electric field satisfies the coupled non linear equations
\bea 
j &=&  j^d(E) {E_x\over E}+\sigma_{H}E_y, \nonumber\\
0&=& -\sigma_{H} E_x +  j^d(E) {E_y\over E}.
\label{curr-HB}
\eea
In the case of large Hall angle, \be
\sigma_{H}>>\sigma_{xx} \equiv  j^d/E,
\label{largeHA}
\ee
and thus $E\approx E_y>> E_x$. This inequality
simplifies the solution of  (\ref{curr-HB}): 
\be
E_y (j)   \approx  { j^d (E)  \over \sigma_{H}}\left(1 + \cO(|\sigma_{xx}|/\sigma_{H})  \right) ,
\ee
where we can drop the terms of order $\sigma_{xx}/\sigma_{H}<<1$. An experimental measurement
of  $E_x(j)$ approximately describes  the non linear function $j^d(E)$, which will be calculated in Eq. (\ref{current-final}).
\be
E_x(j)  \approx   {1\over \sigma_{H}}  j^d\left( {j\over \sigma_{H}}\right).
\ee
The negative conductivity implies  $\bE\cdot \bj <0$. In the ZRS, the  
spontaneous fields $E^{zrs}$ of  (\ref{ZCS}),  are associated with disssipationless flow of  Hall  currents with magnitude 
\be
|j^{zrs}| = \sigma_{H}~E^{zrs}.
\ee
In the ZRS, the currents traverse the system in the $x$ direction  producing very little longitudinal voltage drop and dissipation.
For the  Hall bar geometry of width $W$ (see Fig. \ref{fig:Geometry}),  the resistivity remains vanishingly small upto a  critical current $I^{cr}$ 
given by 
\be
I^{cr} =  W  \sigma_{H} E^{zrs}.
\ee
Thus,  the critical current directly measures the internal ZRS field.

\section{The Microscopic Model}
\label{sec:Model}
We consider a two dimensional electron gas (2DEG)  subject to a perpendicular magnetic field $B\hat{\bz}$, microwave field  $\bE_\omega$,
and a DC  field $\bE_{dc}$.  The standard treatment of electrons in a magnetic field defines  dimensionless Landau level and guiding center coordinates as
 \bea
\bPi &\equiv& {l_B\over \hbar}  \bp+{1\over 2 l_B}\hat{\bz}\times\br,\nonumber\\
\bR &\equiv&   {\br\over 2 l_B}+ {l_B\over \hbar} \hat{\bz}\times\bp  ,
 \label{coordinates}
\eea
where the magnetic length is $l_B=\sqrt{\hbar c/eB}$.
By construction, the two guiding center components commute with the two Landau operators:
\be
\left[  \pi^\alpha ,R^\beta \right]=0,~~~~~\alpha,\beta=x,y.
\label{PiR}
\ee

\begin{figure}[htb]
\begin{center}
\includegraphics[width=9cm,angle=0]{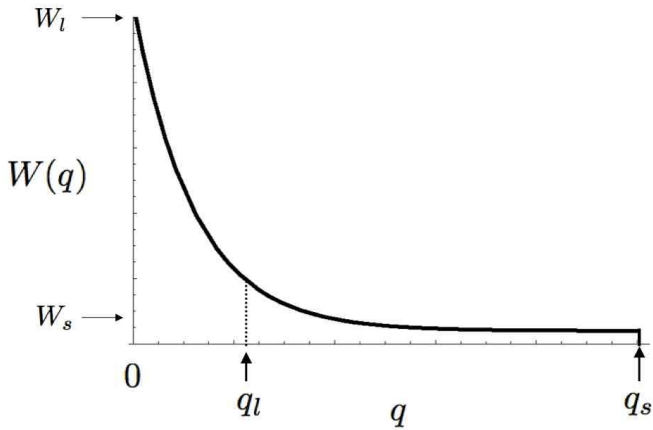}
\caption{The  `Broad-Tail' model of disorder correlations. The wavevector $q_l$ bounds the 
long wavelength spectrum which is incorporated into the disordered Floquet evolution operator (see text).  Most of the spectral weight is below $q_l$, although the higher moments which determine the dissipative current, are dominated by the short wavelength cut-off  $q_s$.}
\label{fig:Wq}
\end{center}
\end{figure}

The microscopic model Hamiltonian is
\be
\cH(t) =  {\hbar \omega_c\over 2} \left| \bPi+\ba(t) \right|^2  + \cV(\br),
\ee
where $\omega_c= eB/(mc)$ is the cyclotron frequency.  The electric fields are represented by the dimensionless gauge field
\be
\ba(t)= {e l_B \over \hbar} \left( \Re  \left[{{{\bf E}_\omega}
\over \omega} e^{-i \omega t}\right] +  {\bf E}_{dc} t \right).
\ee
The disorder potential is described  by a random function of position $\br$,
\be
\cV= {1\over \cA} \sum_\bq V_\bq e^{-i\bq\cdot \br},
\ee
where  $\cA$ is the area of the system.
The Fourier components are random complex numbers obeying $V_\bq=V_{-\bq}^*$, with correlations
\be
\langle V_\bq V_{-\bq'} \rangle =  \cA   W(q) \delta_{\bq,\bq'}.
\label{dis}
\ee
Our {\em  'divide and conquer' } tactic is to split the disorder potential into two components by a dividing wavevector $q_l$.  
We assume that   $W(q)$ is a {\em `broad-tail' distribution}, (see Fig.~\ref{fig:Wq}), i.e.,  most of its  weight is concentrated in a region below some dividing wavevector $q_l$, (say, of the order of $l_B$), but the higher  {\em moments}
are dominated by  much  shorter wavelengths.
For example, we might choose the function
\bea
 W(q) &=& W_l e^{-2q/q_l} + W_s \theta(q_s-q),\nonumber\\
  {q_l\over q_s} &< &{W_s q_s \over W_l q_l} < 1,
 \label{disorder}
 \eea
 such as depicted in Fig.~\ref{fig:Wq}. 

 This model is realistic for high mobility heterostructures, where the offset distance to unscreened charges is $d=1/ q_l$ \cite{Efros}, and $W_s$ is the  intrinsic short range disorder of the 2DEG. The upper cut-off is effectively at some $q_s\le \pi k_F$. 
We do not include at this level any  electron phonon or electron electron interactions.
 
\subsection{Splitting the Disorder}
The main trick is to split  the disorder into two operators  
\be
\cV(\br)= \cV_l (\bR) + \cV_s(\bPi,\bR)
\label{V+V}
\ee
The long wavelength disorder term is {\em defined } as the operator
\be
\cV_l \equiv \sum_{nkk'} V^{n_F}_{kk'} |nk\rangle\langle nk'|=\sum_{n\alpha} \epsilon_\alpha |n\alpha\rangle \langle n\alpha|.
\label{Vl}
\ee
 $n_F$ is the single-spin filling factor at the Fermi energy.
 The Landau level degeneracy is  $n_L=  \cA/(2\pi l_B^2)$,
where $\cA$ is the system area.
Without loss of generality we choose the  basis $|n k\rangle$ on a cylinder of circumference $L_x =\sqrt{\cA}$, where $n$ is the Landau level index, and $k=  2\pi m l_B/ L_x$, where $m = 1, \ldots n_L$. The  state $|k\rangle$ is centered at position $\langle y\rangle =  k l_B $.
The matrix elements of $\cV_l$ are the (identical) block-diagonal matrices (see Appendix 
\ref{App:disorder}):
\be
V^{n_F}_{kk'}= {1\over \cA} \sum_{|\bq|\le q_l} V_\bq e^{- {\tq^2 \over 4}}  L_{n_F}\left({\tq^2\over 2}\right) e^{-i \tq_x {(k+k')\over2}} \delta_{k+\tq_y,k'}.
\label{Vknnp}
\ee
$L_{n}$ is the Laguerre polynomial of order $n$, and $\tq \equiv q l_B$. In (\ref{Vl}), 
$\epsilon_\alpha$ and  $\varphi_\alpha(k)=\langle k|n\alpha\rangle$ are the disordered spectrum and eigenvectors respectively.
 
 $\cV_l$, by construction, has no inter-band
matrix elements, and identical intra-band blocks.
It is therefore a random function of {\em only }the guiding center  
operators $\bR$ (see Eq. (\ref{coordinates}).  Nevertheless, it contains most of the spectral weight of the disorder.


\begin{figure}[htb]
\begin{center}
\includegraphics[width=9cm,angle=0]{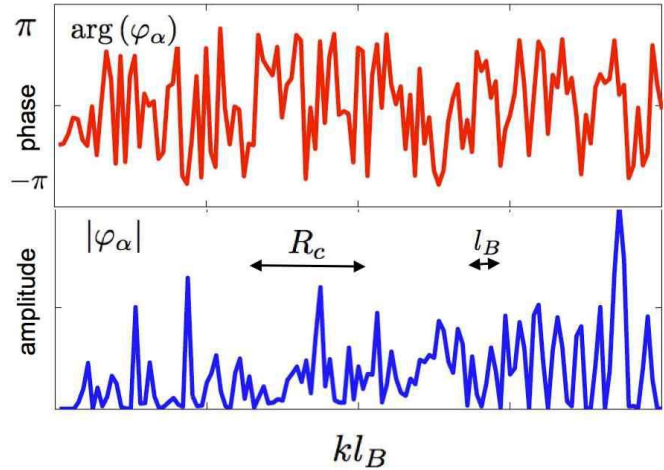}
\caption{A  typical random eigenstate $\varphi_\alpha$, chosen near the broadened Landau  band center.
 The phase (amplitude)  is plotted in the top (bottom) panel.
 $kl_B$  describes the height position on the cyclinder. We use
Landau degeneracy $n_L=128$,  
Fermi level $n_F=15$, and   cut-off $q_l=2/l_B$ for the long range disorder. 
 The cyclotron radius and Landau length are depicted.}
\label{fig:WF}
\end{center}
\end{figure}


\subsection{The Random Matrix Approximation}

We shall now see that at high Landau levels, $V^{n_F}_{kk'}$  is  well represented by a random Hermitian matrix. 
Its eigenstates appear as random extended wavefunctions  as depicted in Fig. \ref{fig:WF}. 

Although   numerical averaging over disorder poses no great difficulty, the statistical properties of $V^{n_F}_{kk'}$  suggest to approximate it  as a Hermitian random matrix
of the Gaussian Unitary Ensemble (GUE).
This Random Matrix Approximation (RMA)  will provide easy  averaging of spectra and correlations over the long wavelength disorder.
This presents another example of the usefulness of Random Matrix Theory \cite{Mehta} for quantum transport  \cite{RMT}.

Fortunately,  the
RMA becomes exact in the limit of a uniform white-noise disorder fluctuations $W(q)=W$, and large filling factor. The matrix elements
of $V^{n_F}_{kk'}$  become  maximally uncorrelated:
\bea
&&\lim_{n_F\to \infty} \langle \cV^{n_F}_{kk'} \cV^{n_F}_{k''k'''} \rangle_{V_\bq}\nonumber\\
&& ~~~~~= {1\over \pi R_c \cA} \sum_{\bq} {W(q) \over \sqrt{q q'}}  e^{i\tilde{q}_x(k-k''')}\nonumber\\
&&~~~~~~~~~~~\times  \delta_{k',k+\tq_y } \delta_{k'',k'''+\tq_y }\nonumber\\
&&  ~~\approx \left({2 \pi l_B^2 \over \cA}\right)  \Gamma_{RMA} \delta_{k,k'''}\delta_{k',k''}\nonumber\\
&&\Gamma_{RMA}^2 = {1\over 2 \pi^2 R_c} \int_0^{\infty} dq W(q) ,
\label{Gamma-GUE}
\eea
where $R_c =  l_B\sqrt{2 n_F}$. 
The density of states is defined as an ensemble average 
\be
\nu(E)=\langle \sum_{n\alpha}\delta(E-E_{n\alpha})\rangle_{\cV_l}.
\ee
For a GUE ensemble, the density of states of each Landau band is expected to be semi-elliptical. 
In the limit of narrow Landau levels, the  SCBA also produces semi-elliptical density of states. 
Moreover, the  SCBA \cite{AU,RS} for the Landau level broadening (sometimes called in the literature `quasiparticle scattering rate' $\hbar/\tau_s$) 
 equals $\Gamma_{RMA}$,  within the assumptions of Eq. (\ref{disorder}).

Here, however, we are interested in  long range disorder where $q_l<< k_F <\infty$. This implies that  the matrix elements of
$V^{n_F}_{kk'} $ are correlated within a distance
 $\Delta k \le   1/(l_B  q_l)$, and that the density of states at  the tails will not be well described by the RMA's semi-elliptical form.

\begin{figure}[htb]
\begin{center}
\includegraphics[width=8.5cm,angle=0]{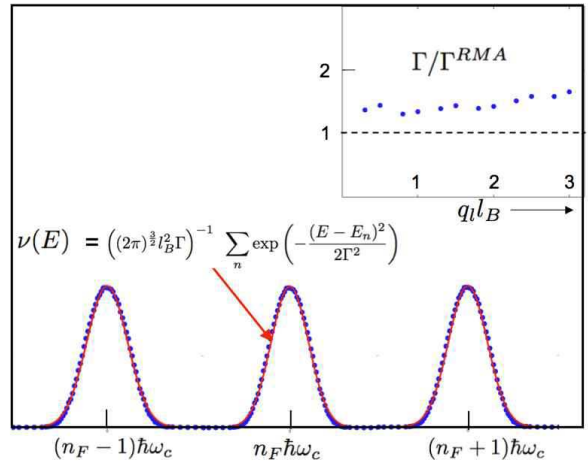}
\caption{Numerical averaging of 500 disorder potentials, for the density of states $\rho(\epsilon)$ of $V^{n_F}_{kk'}$, using
 $W_l=l_B^2$, $n_F=80$,  Landau level degeneracy
$n_L=512$ and $q_l=2/l_B$. The fit is to a Gaussian of deviation $\Gamma$. Inset: dots are the numerical values of $\Gamma$
for different cut-off wavevectors $q_l$, divided by  the Random Matrix Approximation $\Gamma^{RMA}$, Eq.~(\ref{Gamma-GUE}).} 
\label{fig:DOS}
\end{center}
\end{figure}

In the regime $q_l l_B \in [0.3,5]$ the  numerical density of states, averaged over a Gaussian distribution of disorder potentials  $\{ V_\bq \}$, can be fitted  to 
a sum of Gaussians:
\bea
\nu(E) \simeq {1\over (2\pi)^{3\over 2} l_B^2 \Gamma } \sum_n e^{ -{(E-n\hbar\omega_c)^2\over 2\Gamma^2 }}.\nonumber\\
\label{dos}
\eea
As shown in  Fig.\ref{fig:DOS} the Gaussian fit is excellent for  $q_l l_B =2$. The fitted  Landau level width $\Gamma$ is also
quite close to the RMA result,
\be
\Gamma\approx 1.3 \Gamma^{RMA}.
\label{GammaRMA}
\ee
The RMA for the  eigenfunction correlations, which  determine the  structure factor,  will be
investigated later in Section 
\ref{sec:RMA-SF}.

\section{Floquet Transformation of  Time Dependent Fields}
 \label{sec:Floquet}
 \subsection{Clean Landau Levels}

The clean Hamiltonian,  driven by a time dependent vector potential is
\be
\cH_0(t)= {\hbar \omega_c\over 2} \left| \bPi+\ba(t) \right|^2  .
\label{H0}
\ee 

The Schr\"{o}dinger equation for the evolution operator
\be
i\hbar \partial_t U_F(t,t_0)=\cH_0(t) U_F(t,t_0),
\ee
is solved in  Appendix \ref{App:Floquet} and is given by
\bea
U_F(t,t_0) &=&   e^{-i \bxi(t) \cdot\bPi }~ e^{-i{\bar\cH}_0 (t-t_0)}e^{+i \bxi(t_0) \cdot\bPi },\nonumber\\
\bar{\cH}_0 (t) &=&  \sum_{nk}  \hbar n \omega_c   |nk\rangle\langle n k |~,
\label{UF}
\eea
where the Floquet Hamiltonian  $\bar{\cH}_0$ is identical to the time independent Hamiltonian. This identity,  (called Kohn's theorem)  
also holds in the presence of arbitrary two-body interactions $\sum_{i<j} V(\br_i-\br_j)$ between electrons  \cite{Kohn}.

The vector  fields $\bxi(t)= \bxi_\omega+\bxi_0$ are solutions of coupled linear equations. Their explicit form, given in   Appendix \ref{App:Floquet},   is well known \cite{Ryzhii}:
\begin{eqnarray}
\xi_x^\omega  &=& - {e l_B \omega_c \over \hbar \omega}  \Re \left[{{\omega_c  E^y_\omega + i \omega  E_\omega^x}
\over {\omega_c^2-\omega^2}} e^{-i \omega t}\right] , \nonumber\\
\xi_y^\omega  &=& {e l_B \omega_c \over \hbar \omega}  \Re \left[{{\omega_c  E^x_\omega - i \omega  E_\omega^y}
\over {\omega_c^2-\omega^2}} e^{-i \omega t}\right] ,\nonumber\\
\bxi_0(t)&=&   t {e l_B \over\hbar  } {\bE}_{dc}\times\hat{\bz} - {e l_B\over\hbar\omega_c  }{\bE}_{dc}.
\label{etaxi}
\end{eqnarray}

\subsection{Floquet Transformation With  Disorder}
In Eq. (\ref{Vl})  we have constructed the long range disorder operator  $\cV_l$ to a function only of $\bR$.
Thus, Eq.~(\ref{PiR}) implies the trivial,  {\em yet very useful}, commutation relation
\be
[\cV_l,\bPi]=0,
\ee
which we exploit to write the time dependent disordered Hamiltonian as
\be
\cH_d(t)= \cH_0(\bPi,\ba(t)) + \cV_l[\bR].
\label{Hd}
\ee 
Now we can construct the {\em disordered Floquet
evolution operator} $U_d$, which obeys the evolution equation 
\be
i\hbar \partial_t U_d(t,t_0)=\cH_d(t) U_d(t,t_0).
\ee
Since the Floquet operators $e^{i\bxi(t)\cdot \bPi}$ commute with $\cV_l[\bR]$,  
the solution is simply
\bea
U_d(t,t_0) &=&   e^{-i \bxi(t) \cdot\bPi }~ e^{-i{\bar\cH}_d (t-t_0)}e^{+i \bxi(t_0) \cdot\bPi },\nonumber\\
\bar{\cH}_d  &=&  \sum_{n\alpha}  \left(\hbar n \omega_c+\epsilon_\alpha\right)  |n\alpha\rangle\langle n\alpha|~,
\label{Ud}
\eea
with the same vector function  $\bxi (t)$ as for the clean Landau level problem (\ref{etaxi})!
For $\bE_{dc}$=0,
the Floquet states $|n\alpha(t)\rangle =  U^d(t,t_0) |n,\alpha\rangle$ 
return to themselves (up to a  phase) at every integer period. 

Expression (\ref{Ud}) implies that the Floquet quasi-energies (the spectrum of the
transformed Hamiltonian $\bar{\cH}_d$) is the {\em disorder broadened} Landau  levels. The equivalence between stationary and quasi-stationary states
with time dependent electric fields,  is a special property shared by harmonic oscillators, Landau levels, and our $\cH_d$.

Using the last equation in (\ref{Floq-trans}), we can also explicitly express  the  time dependent  Fourier operator 
as
\be
e^{i\bq\cdot\br(t)}=  e^{{i\over \hbar} \bar{\cH}_d t } e^{i\bq\cdot\br} e^{-{i\over \hbar} \bar{\cH}_d t }
~e^{i l_B \bq\cdot\bxi(t)},
\label{eiqrt1}
\ee
where we can use the explicit solution (\ref{etaxi}) to perform the Bessel expansion,
\be
 e^{i l_B \bq\cdot\bxi(t)} =  e^{-i\psi} \sum_{\nu= \infty}^\infty i^\nu J_\nu \left(\tq \Delta_{\omega\hbq}\right)  e^{it ( \nu \omega + (e l_B^2 / \hbar) \bq \times \bE_{dc}\cdot \hat{\bz})  } ,
\label{eiqrt2}
\ee
where $\psi$ is an unimportant phase, and 
$J_\nu$ is the Bessel function of order $\nu$. The quantity
\be
\Delta_{\omega\hbq} (\bE_\omega) \equiv   \mbox{max}_t \left[ \hat{\bq}\cdot\bxi(t) \right]  
\label{Delta1}
\ee 
is proportional to  the 
microwave field strength.
For positive circular polarized light of field $\bE_\omega=E^+_\omega(\hat{\bx}+i\hat{\by})$, 
$\bxi$ and $\Delta_{\omega\hbq}$ diverge  at the cyclotron resonance 
as $|\omega-\omega_c|^{-1}$. For the opposite polarization, $\Delta_{\omega\hbq}$ is continuous at resonance. 

\section{Calculation Of  The  Current}
\label{sec:current}
The DC current density  is  given by the full evolution operator $U$ as 
\be
\bj =  {e  \hbar\over  \cA m l_B} \sum_{n\alpha}\rho_{n\alpha} \langle n\alpha|  
 \overline{U^\dagger(t,t_0)\bPi U(t,t_0)  +\ba(t)}  |n\alpha\rangle ,
 \label{bjexact}
\ee
where $\rho_{n\alpha}$ is the electron density operator, defined in the Floquet 
basis $|n,\alpha\rangle$.

The notation
$\overline{F(t,t_0)}$ implies averaging the end-point times $t,t_0$ over many periods of $2 \pi/\omega$.

\subsection{Zeroth Order Theory}
The zeroth order DC current density is given by setting $U \to  U_d$ in (\ref{bjexact}).  Using (\ref{Floq-trans}) and (\ref{etaxi}), it is easy to obtain
\bea
\bj^{(0)} &=& \left( \bxi_0\times{\hat\bz} +\ba_{dc}  \right)  {e  \hbar\over  m l_B\cA  } \left(\sum_{n\alpha}\rho_{n\alpha}\right)\nonumber\\
&=&     {e^2 n_e \over  m \omega_c}~\hat{\bz} \times  \bE_{dc},
\label{bj0}
\eea
where the result holds for {\em any distribution}   $\rho_{n\alpha}$ (not necessarily Fermi-Dirac) which sums up to the full electron density $n_e$.

It is interesting that although $\bar{\cH}_d$ has random spectrum and wavefunctions, its
current  density (\ref{bj0}) produces no dissipation. In addition we see that the Hall conductivity,
is precisely given by the classical  value,
\be
\sigma_H= {e^2 n_e \over  m \omega_c}= n_F {e^2\over h},
\label{sigmaH}
\ee
although for $\cV_l\ne 0$, our  system is by no means Galilean invariant. Incidentally, this result
(\ref{bj0})  applies in the presence of  arbitrary electron-electron interactions which may
be  added to $\cH_d(t)$.

We iterate  that Eq.  (\ref{bj0})  applies only to the zeroth order Hamiltonian defined in Eq. (\ref{Hd}). 
Dissipative effects enter at second order   
$\cO(\cV_s^2)$.

\subsection{Second Order Dissipative Current}
\label{App:current}
The leading order dc current, requires expanding the momentum operator to second order in  $\cV_s$.

The interaction representation is defined as,
\be
\cO^d(t,t_0) \equiv U^\dagger_d(t,t_0) \cO U_d(t,t_0).
\ee

The zeroth order ('clean limit') Landau operator  (in complex notation $\tilde\bPi=\bPi^x+i\bPi^y$) transforms as
\be
{\tilde\bPi}^{(d)}(t,t_0) = e^{i\omega_c (t-t_0)}  {\tilde \bPi} ~  - \gamma(t,t_0),
\ee
where $\gamma$ is an unimportant c-number function.

The full evolution operator $U$ is expanded to second order in $\cV_s$, 
\bea
U(t,t_0)&=&  U_d(t,t_0) \left(1+{- i\over \hbar}\int_{t_0}^t dt'  \cV_s(t',t_0)\right.  \nonumber\\
&& \left.  +{(-i)^2\over \hbar^2} \int_{t_0} ^t \!dt'  \int_{t_0}^{t'} \!dt''  \cV^d_{s}(t,t_0) \cV^d_{s}(t'',t_0) +\ldots\right),\nonumber\\
\eea
 which yields the second order correction to the Landau operators as
\bea
&&{\tilde\bPi}^{(2)}(t,t_0) = {1\over\hbar^2}  \int_0^t\! dt' \int_0^{t'} \! dt''  e^{i\omega_c(t-t_0)}\nonumber\\
 &&~~~~~~\times\left[\left[  {\tilde\bPi}, \cV^{(d)}_{s}(t',t_0)\right],\cV^{(d)}_{s}(t'',t_0)\right] \nonumber\\
&&=  {1\over m l_B \hbar A^2} \sum_{\bq\bq'} V_\bq V_{-\bq'}  (q_x+iq_y)\int_0^t \!dt' \int_0^{t'}\! dt''\nonumber\\
&&~~~~~e^{i\omega_c(t-t')} \left[ e^{i\bq\cdot\br^{(d)}(t',t_0)},
 e^{-i\bq'\cdot\br^{(d)}(t'',t_0)}\right].\nonumber\\
 \label{bv}
 \eea

The leading order dissipative current is obtained by the disorder averaged expression,
\be
{\bj}^d  
\simeq  {e\hbar \over \cA m l_B}  \left<\sum_{n\alpha} f_{n\alpha}  \left< \langle n\alpha |\overline{\bPi^{(2)}(t,t_0)} |n\alpha\rangle \right>_{\cV_s}  \right>_{\cV_l} .
\label{bjr} 
\ee 
Here we use the Fermi-Dirac distribution
\be
f_{n,\alpha} = \left( 1 + e^{-E_{n\alpha}/T} \right)^{-1} .
\ee
Corrections to the distribution function enter at higher order in $\cV_s$, see discussion in Section \ref{sec:Discussion}, item 2.
Eq. (\ref{bjr}) ignores correlations between $\cV_l$ and $\cV_s$, which exist  in the range
$q<q_l$.
The resulting error in the current, which is dominated by
large wavevectors, is relatively suppressed by powers of $q_l^2/q_s^2$.

 Using the matrix elements of the time dependent Fourier operator (\ref{eiqrt1})  we obtain,
\bea
&&\langle n\alpha| e^{i\bq\cdot\br^{(d)}(t,t_0)}|m\beta \rangle= e^{{i\over \hbar}(E_{n\alpha}-E_{m{\beta}})(t-t_0) } D_{nm}(\bq) \nonumber\\  
&&\quad \times \left( \sum_k e^{iq_x kl_B} \varphi_\alpha^*(k)\varphi_{\beta}(k+q_y l_B)\right)  \nonumber\\
&&\quad \times \sum_{\nu= -\infty}^\infty J_\nu \left(\tq \Delta_{\omega\hbq} \right)  e^{it ( \nu \omega + (e l_B^2 / \hbar) \bq \times \bE_{dc}\cdot \hat{\bz})  } 
\nonumber\\
\label{eiqrt3}
\eea
 Averaging (\ref{bv}) over $\{  V_\bq \}$  keeping $\cV_l$ fixed,  yields  \bea
&&\langle n \alpha| \tilde{\bPi}^{(2)}( t,t_0) |n \alpha\rangle_{\cV_s} ={1\over \hbar \pi R_c  A}\sum_{\bq m} {q_x+iq_y\over q} \nonumber\\
&&~~\quad\quad\quad\quad\quad\quad \quad\quad\quad W(q)  \left(1 -\delta_{nm} \theta_{q<q_l}\right)      I_{nm}(t,t_0) ,\nonumber\\
&& I_{nm} (t,t_0)= \sum_{\alpha'}\int_{0}^{t-t_0} \! dt' e^{i\omega_c(t-t')}\nonumber\\
&&~~~ \int_{0}^{t'} \! dt'' {1\over 2} \sum_{\pm} \pm e^{\pm i(E_{n\alpha}-E_{m\alpha'})(t'-t'') }
\nonumber\\
&&~~\quad\quad\quad \times e^{i\bq\cdot (\bxi(t'+t_0)-\bxi(t''+t_0))}  \langle e^{i\tilde{\bq}\cdot \bR } (t) e^{-i\tilde{\bq}\cdot \bR}(t') \rangle_{\cV_l} .\nonumber\\
\label{v2nn}
\eea
We can henceforth neglect the terms $\left(-\delta_{nm}\theta_{q<q_l}\right)$
in (\ref{v2nn}). The corrections contribute only to the low wave vector intergration, and are therefore suppressed by factors of $(q_l/q_s) <<1$.

Integrating (\ref{bjr}) over time, we obtain the 
the full current expression,
\bea
\bj^d    &\simeq& {e \over  m R_c \omega_c}\int^{\pi k_F} {d^2 q\over (2\pi)^2 }  ({\hat \bz}\times \bq)  
\sum_{nm} {W_{nm}^{res}(q)\over q}
\nonumber\\
&& \times \int d\epsilon d\epsilon'
\left(f_{n,\epsilon}-f_{m,\epsilon'}  \right) K_2(q,\epsilon,\epsilon' ) \sum_\nu |J_\nu( \tq \Delta_{\omega\hbq})|^2\nonumber\\
&&\times\delta(E_{m\epsilon'}-E_{n\epsilon}-\hbar \nu \omega-e l_B^2   {\hat \bz}\times \bq\cdot \bE_{dc}).\nonumber\\
\label{current-final}
\eea
$K_2$ is the disorder averaged intra-band structure factor,
\bea
&&K_2(q,\epsilon, \epsilon') =  
\int {dt dt'\over (2\pi \hbar)^2} e^{{i\over \hbar}(\epsilon t-\epsilon't')} \langle e^{i\tilde{\bq}\cdot \bR } (t) e^{-i\tilde{\bq}\cdot \bR}(t') \rangle_{\cV_l}
\nonumber\\
&&= {1\over \cA} \sum_{k k'\alpha\ne \beta} e^{i \tq_x (k-k') }  \nonumber\\
&&  ~~~~~~~~ \times\left< \varphi^*_\alpha(k)\varphi_\alpha(k')\varphi^*_\beta (k+\tq_y)\varphi_\beta (k'+\tq_y) \right.\nonumber\\
&& \quad\quad\quad\quad\quad\quad\quad \quad~~~\times \left.  \delta(\epsilon-\epsilon_\alpha)
\delta(\epsilon'-\epsilon_\beta)\right>_{\cV_l}. \nonumber\\
\label{Kernel}
\eea
Note that by the symmetry of   $\int d^2q$ integration,  the second order current is
parallel to the DC field, and purely dissipative.  
Hence to second order,  {\em  the Hall conductivity remains equal to the classical value} $\sigma_H$ of (\ref{sigmaH}).

\subsection{ Simplification at $T>\hbar \omega_c$}
Many of the relevant experiments are  carried out  at temperatures $T > \hbar\omega_c$,
where the Shubnikov-de Hass oscillations are thermally smeared. 
Using the energy-conserving delta function, we approximate  the 
difference of Fermi functions as
\be
  f_{n,\epsilon}-f_{m,\epsilon'} \approx    -\Theta\left( T- |\Delta E_\nu| \right)~\Delta E_{\nu}{\partial f(E_{n\epsilon})\over \partial E} ,   \ee
where $\Delta E_\nu$ is the transition energy 
  \be
  \Delta E_{\nu} = \hbar   \nu  \omega + e l_B^2   {\hat \bz}\times \bq\cdot \bE_{dc} .
  \label{Delta-nu}
  \ee
Defining the single frequency structure factor  $K_1$ as
\be
K_1(q,\omega)=  \int d\epsilon K_2(q,\epsilon,\epsilon+\hbar\omega),
\label{K1}
\ee
the current expression simplifies to 
 \bea
j^d  &\simeq& {e  \over  4 \pi^2 m v_F }\int^{\pi k_F} \!d^2q ~ {W(q)  \over q} ({\hat \bz}\times \bq)\nonumber\\
&& ~\times\sum_{\nu, m}^{{|\Delta E_{\nu}|\le T})}  
\left( \hbar \nu \omega + e l_B^2   {\hat \bz}\times \bq\cdot \bE_{dc}\right) 
|J_\nu( \tq \Delta_{\omega\hbq})|^2 \nonumber\\
&&~~~~~~~~\times K_1\left( q, m \omega_c +  \nu \omega + e l_B^2   {\hat \bz}\times \bq\cdot \bE_{dc}/\hbar\right) .\nonumber\\
\label{current-simple}
\eea

\begin{figure}[htb]
\begin{center}
\includegraphics[width=8cm,angle=0]{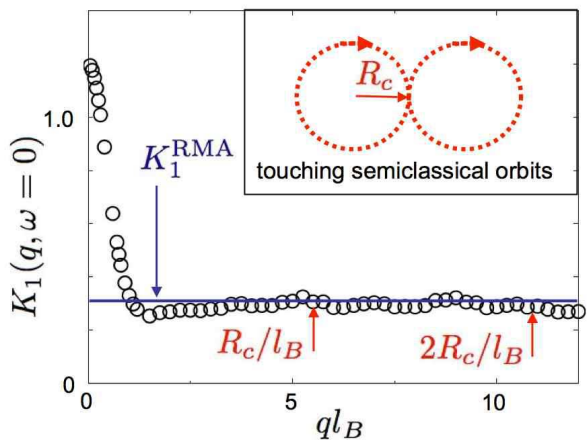}
\caption{The static structure factor $K_1(q,\omega=0)$ evaluated numerically for  $n_L=128$,   $n_F=15$, and $q_l=1/l_B^2$. 
The exact correlator has a peak at zero waavevector of width  $q_l$, and agrees with the (wavevector independent) RMA  at higher wavevectors. The Inset depicts {\em classica}l cyclotron orbits, which strongly overlap when  separated by $2R_c$.
In contrast,  $K_1$ shows no signature of enhanced scattering at  wavevectors corresponding to $2R_c$. } 
\label{fig:K1-stat}
\end{center}
\end{figure}

\subsection{The RMA For The Structure Factors}
\label{sec:RMA-SF}

Within the RMA, the Hamiltonian of each Landau level is taken as  a member of 
the Gaussian Unitary Ensemble (GUE).
 Thus, by the RMA,  the wavefunctions  $\varphi_\alpha(k)$ are  normalized complex random vectors, and the spectral correlations
are completely uncorrelated beyond a few level spacings \cite{Mehta}.  From random matrix theory, one has
\bea
\left< \varphi^*_\alpha(k) \varphi_\beta(k')\right>_{\cV}\nonumber &=&\delta_{\alpha\beta}\delta_{kk'},   \nonumber\\
{1\over \cA^2} \left<\sum_{\alpha\beta}\delta(\epsilon-\epsilon_\alpha) \delta(\epsilon-\epsilon_\alpha)\right>_{\cV}&=& 2\pi l_B^2 \rho(\epsilon)\rho(\epsilon'),
\nonumber\\
\eea
for $\epsilon-\epsilon' >> {\Gamma\over n_L}$.
The RMA approximates  the dynamical structure factor  Eq. (\ref{Kernel}) as 
\be
K_2^{RMA}(\epsilon,\epsilon') =  {1\over (2\pi)^2 l_B^2 \Gamma^2} \exp \left( - {\epsilon^2+ (\epsilon')^2  \over 2\Gamma^2} \right) ,
\label{K-GUE}
\ee
where $\Gamma=\Gamma^{RMA}$ is the width of the single Landau level density of states (\ref{Gamma-GUE}).   

\begin{figure}[htb]
\begin{center}
\includegraphics[width=8.5cm,angle=0]{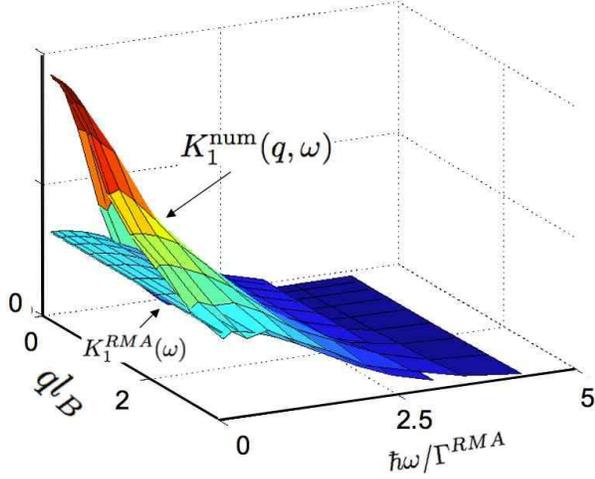}
\caption{The dynamical structure factor.  $K^{\rm num}_1(q,\omega)$ is evaluated numerically, for $n_L=64$,  $n_F=80$
and $q_l=1/l_B$.  The Random Matrix Approximation  $K_1^{RMA}$ is given in Eq. (\ref{K1RMA}). Good agreement is found for $q> 3q_l$,
at all frequencies. } 
\label{fig:K1-dyn}
\end{center}
\end{figure}

The  RMA for the single frequency  structure factor is 
\bea
K_1^{RMA}(\omega)&=& \int d\epsilon K_2^{RMA}(\epsilon,\epsilon+\hbar\omega)\nonumber\\
&=& {  \sqrt{\pi} \over (2\pi)^2 l_B^2  \Gamma} \exp\left( - {\hbar^2\omega^2 \over 4\Gamma^2}\right) .
\label{K1RMA}
\eea 
We have  computed  the structure factor $K_1(q,\omega)$ numerically for a finite Landau level degeneracy on a cylinder. As depicted in
Figs. \ref{fig:K1-stat} and  \ref{fig:K1-dyn},  the RMA works well for   $q>>q_l$, where
$\Gamma$ is given by  (\ref{GammaRMA}).
This finding confirms our expectation that the  RMA  is applicable at large wavevectors,  since correlations between matrix elements of $V^{n_F}_{kk'}$,  decay beyond the wavevector scale $\tilde{q}_l$.
In contrast, at  low wavevectors   $q\le 3 q_l$, there is a peak in $K_1$, which is absent in the RMA. It contributes negligibly however, to the overall current. We note that
there is no  signature of enhanced scattering at wavevectors corresponding to 
the classsical cyclotron diamater $2R_c$, see discussion in section \ref{sec:Discussion}, item 6.

\section{Analytical Results}
\label{sec:Results}
The full non linear current $\bj^d (\bE_{dc},\bE_\omega,\omega,B,T,n_e)$ can  be computed from Eq. (\ref{current-final}),  by numerically averaging
$K_1$ over disorder.
It is useful however to use the RMA and obtain analytical results. These  allow a better understanding of the important characteristic
energy and field scales of the non linear current.

The first order of   business,  is to extract  the dark linear conductivity from Eqs. (\ref{current-simple}) and  (\ref{K1RMA}). We obtain:
\be
\sigma^{\rm dark}_{xx} ={n_F e^2 \over h} \left( {\hbar \over   2  \pi^{1\over 2}\tau_{tr} \Gamma} \right) ,
\label{dark}
\ee
Not surprisingly, this  result agrees with the SCBA \cite{comm-dark}, where
the {\em transport scattering rate} is defined in the standard   way 
\cite{Dietel}:
 \be
{1/\tau_{tr}} \equiv( 2 \pi \hbar^2 v_F k_F^2 )^{-1} \int_0^{q_s} dq q^2 W(q) .
\label{tau-tr}
\ee

Using the RMA for $K_1$, given by (\ref{K1RMA}),   the  current  can be simplified into an analytical expression:
\bea
j^d  &=&  \sigma_{xx}^{\rm dark} E_{\Gamma}
\Bigg(\varepsilon_{dc} F(\varepsilon_{dc},\varepsilon_{\omega},\omega  ) \nonumber\\
&&~~~~~~~~~~~~+\left({\hbar\omega\over \Gamma}\right) G(\varepsilon_{dc},\varepsilon_{\omega}, \omega) \Bigg).
\label{jx-asympt}
\eea

The dimensionless fields  are defined as,
\bea 
\varepsilon_{dc} &\equiv&  E/E_\Gamma,\nonumber\\
\varepsilon_{\omega}  &\equiv& \Delta_{\omega\hbq_s}  q_s l_B.
\label{dim-var}
\eea
The characteristic 
electric field scale  
\be
E_\Gamma=\Gamma/ (el_B^2q_s),
\label{EG1}
\ee
defines the lowest scale of non-linearity in $j^d$.

At $|\bE|\approx E_\Gamma$, resonant intraband transitions are enabled 
over a distance of $l_B^2 q_s$.   
 
For a white noise short range disorder $W(q) = W_s \theta_{q<q_s}$, the dimensionless functions $F,G$ in (\ref{jx-asympt}) are explicitly evaluated as
\bea
F&=& {3\over \pi  }\sum_{m\nu}^{{|\Delta E_{\nu}|\le T}} \int_{r\le 1}  \! d^2 r  {y^2\over r}    |J_\nu( r \varepsilon_\omega )|^2\nonumber\\
&& ~~~~~~~~~~~~~~~~~\times e^{-{(y \varepsilon_{dc}+ {\hbar(\nu\omega-m\omega_c)/ \Gamma})^2/4}},
\nonumber\\
G &=&  {3\over \pi  }\sum_{m\nu}^{|\Delta E_{\nu}|\le T}  \nu \int_{r\le 1}  \! d^2 r  {y\over r}   |J_\nu( r \varepsilon_\omega )|^2\nonumber\\
&& ~~~~~~~~~~~~~~~~~\times e^{-{(y \varepsilon_{dc}+ {\hbar(\nu\omega-m\omega_c)/ \Gamma})^2/4}},\nonumber\\
\label{g0f1}
\eea
where the sums are limited by  the temperature condition on the transition energies  (\ref{Delta-nu}).

\begin{figure}[htb]
\begin{center}
\includegraphics[width=9cm,angle=0]{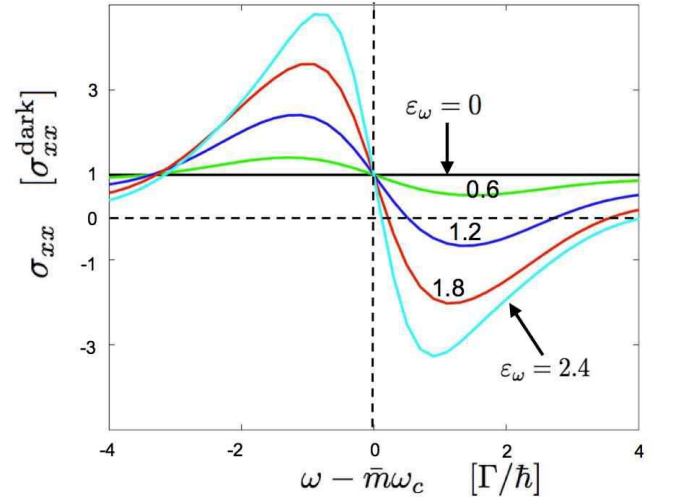}
\caption{MIRO effect. The frequency dependent conductivity is plotted against  detuning frequency near the $\bar{m}^{\rm th}$ resonance.
A set of dimensionless microwave field strengths $\varepsilon_\omega$ are given.  
The broadening ratio (\ref{R}), is chosen at  $R =10$. Temperature is $T=7\hbar \omega_c$.  
The regions of negative conductivity are unstable toward the Zero Resistance State.} 
\label{fig:sigma-dom}
\end{center}
\end{figure}

\subsection{MIRO:  Photo-conductivity Oscillations}
 $G$ describes  the  photo-current contributions.  
Near a particular primary $(\nu=1)$  resonance at some interband interval  $\bar{m}\ge 1$, 
the dimensionless detuning frequency is  defined
as
\be
\delta_\omega ={ \hbar \over \Gamma} (\omega-\bar{m}\omega_c) .
\label{delta-om}
\ee
The MIRO effect, arising  from the function $G$, changes sign at the resonances as
\be
G\propto -\delta_\omega .
\ee
The magnitude of MIRO  is controlled by the 'figure of merit', the  Landau level broadening ratio
 \be
R \equiv {\hbar \omega_c\over \Gamma} = \left( B{2\pi^2 \hbar^2 v_F e\over W_l q_l m c}\right)^{\half} .
\label{R}
 \ee   
For $R>>1$ the total current  at $\delta_\omega>0$,  can turn negative and produce
the Zero Resistance State. This ratio increases as $\sqrt{B}$. Our approximations, however, break down  at low filling factors. 

In Fig.~\ref{fig:sigma-dom}, we plot
a low field conductivity oscillation  as a function of detuning frequency. We choose $R=10$ as the broadening ratio, and  a set of microwave radiation field strengths $\varepsilon_\omega$.
 At temperature $T=7\hbar\omega_c$, we note that the  conductivity at  $\delta_\omega=0$,  is independent of  radiation power.
 The regions of negative conductivity are unstable to   formation of a ZRS.

\subsection{HIRO: The Dark Non-Linear Current}
\label{sec:HIRO}
\begin{figure}[htb]
\begin{center}
\includegraphics[width=8.5cm,angle=0]{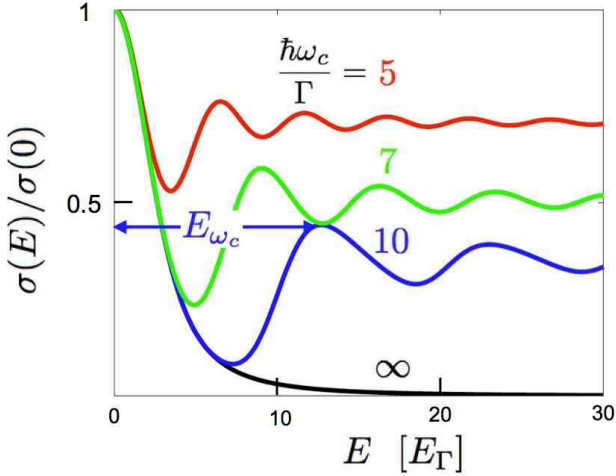}
\caption{The HIRO effect. Non linear current oscillations at zero radiation, as a  function of at  DC electric field, for a set of broadening ratios $R$. 
The field scale is $E_\Gamma$ is defined in (\ref{EG1}). The oscillations periodicity is determined by $E_{\omega_c}=R~E_\Gamma$.}\label{fig:hiro}
\end{center}
\end{figure}

 In the absence of radiation,  $G$ vanishes, and
the non linear conductivity in  (\ref{jx-asympt}) is  $\sigma^{\rm dark}_{xx} F$.
In Fig. \ref{fig:hiro}  we plot $F(E)$ for different values of $\hbar\omega_c/\Gamma$.

At weak fields of order $E_\Gamma$,  we see a Gaussian decrease of conductivity
due to diminishing phase space for intra-Landau bands scattering. Caution has to be exercised in trying to fit the low
conductivity regime with Eq. (\ref{current-final}). The far tails of the density of states require multiple  scattering theory, 
which goes far beyond second order  in  $\cV_s$ \cite{Fogler}.

At larger fields the conductivity has secondary maxima at 
\bea
|\bE^{max}| &=& (m+\delta) E_{\omega_c},~~~~\delta\approx 0.25, m=1,2,\ldots\nonumber\\
E_{\omega_c} &=&    { \hbar\omega_c \over  e l_B^2 q_s}=R~E_\Gamma ,
\label{E-hiro}\eea
which corresponds to   inter-Landau level scattering  over the length scale $q_s l_B^2$.  For  $q_s > \pi k_F$, by  (\ref{asympt}),  
one must use  $q_s \to k_F \pi$.

\subsection{The Non Linear Photocurrent and ZRS Fields}

\begin{figure}[htb]
\begin{center}
\includegraphics[width=8cm,angle=0]{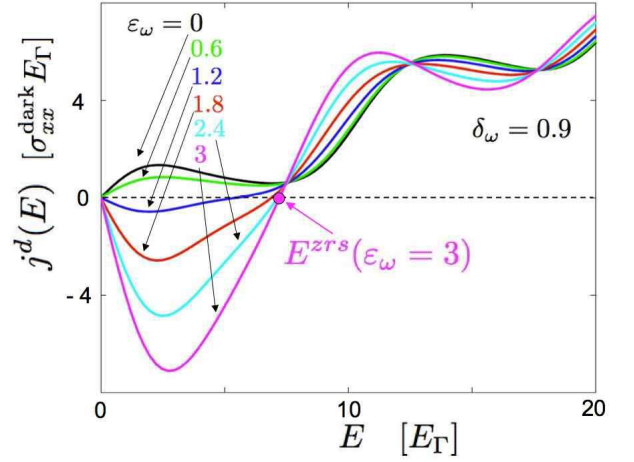}
\caption{Non linear current-field characteristics for differrent levels of microwave field  $\varepsilon_\omega$. The field scale is $E_\Gamma$ (\ref{EG1}). Temperature is chosen at $T=7\hbar\omega_c$, and the broadening ratio is $R=10$. 
 $\delta_\omega$,, defined in (\ref{delta-om}), is fixed at 0.9.  Negative current regimes are unstable to
a ZRS with spontaneous electric fields of magnitude  $E^{zrs}$.} 
\label{fig:JE}
\end{center}
\end{figure}
\begin{figure}[htb]
\begin{center}
\includegraphics[width=8cm,angle=0]{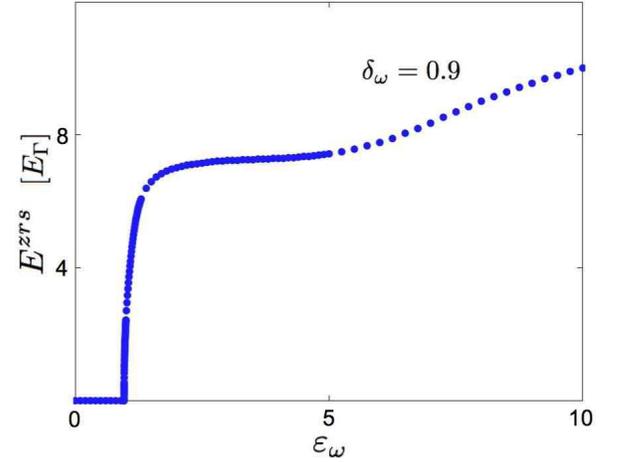}
\caption{The  ZRS spontaneous field. $E^{zrs}$ is given by the zero of $j^d(E)$ as a function of dimensionless microwave field $\varepsilon_\omega$,
at fixed  detuning frequency $\delta_\omega$. A continuous (second order) dynamical   transition occurs at a threshold microwave field.} 
\label{fig:Ezrs}
\end{center}
\end{figure}

In Fig.~\ref{fig:JE}, we plot the fully irradiated non linear current at different microwave fields, at one fixed optimal detuning of
$\delta_\omega=0.9$.  For weak radiation,  the HIRO oscillations are observed, and their phase shifts  as the microwave power increases. 
At stronger radiation, the negative conductivity  regions  are unstable toward the ZRS.
The ZRS creates spontaneous electric fields of magnitude $E^{zrs}$, where the dissipative current vanishes $j^d(E^{zrs})=0$ \cite{Andreev,A1}.
In Fig. \ref{fig:Ezrs}, we plot the dependence of $E^{zrs}(\varepsilon_\omega)$ at fixed detuning frequency. The ZRS exhibits  
a continuous dynamical phase transition as a function of micorwave power, which was previously predicted on phenomenological grounds,  by 
Alicea {\em et  al.}  \cite{Balents}.  

\section{Discussion}
\label{sec:Discussion}

We emphasize several issues regarding  our results.
\begin{enumerate}
\item Our analysis is limited to second order in the short range disorder $\cV_s$, which is controlled by the smallness of 
$ W_s q_s/(W_l q_l) < 1 $. In physical terms, this is equivalent to the well known 'small angle scattering' limit 
where $\hbar/(\tau_{tr}\Gamma)<1$,
with $\Gamma$ and $\tau_{tr}$  defined by Eqs. (\ref{GammaRMA}) and (\ref{tau-tr}) respectively.
 Small angle scattering
is used to justify the SCBA \cite{RS}.  
By  Eqs.~(\ref{sigmaH}) and (\ref{dark}),   the small angle scattering parameter can be {\em experimentally}
determined by the Hall angle $\theta_H$,
\be
\mbox{cotan}(\theta_H) = {\sigma_{xx}^{\rm dark} \over \sigma_H} =  \left( {\hbar \over   2  \pi^{1\over 2}\tau_{tr} \Gamma} \right) .
\ee
By Eq. (\ref{largeHA}),  a large Hall angle $\theta_H\approx \pi/2$ is also the limit which effectively separates the effects of dissipative and Hall fields in the Hall bar geometry \cite{hiro1A,hiro1B}.

\item Our  photo-conductivity supports  the Displacement Photocurrent mechanism \cite{Ryzhii, Durst}.
We have  assumed here, that  electron-electron interactions are relatively weak for  well resolved  high Landau levels.
However, we also implicitly  assume 
that electron phonon interactions are strong enough to produce an inelastic scattering time of the order
\be
\tau_{in} \approx \tau_{tr}.
\ee
Hence, to the leading   order in $1/\tau_{tr}$,  we do not need to correct  the  Fermi Dirac distribution.
We do   not rule out that DF mechanism \cite{Dmitriev} may yield sizeable corrections,
for the case  $\tau_{in} >  \tau_{tr} $.
 
 \item We have assumed a model of weak disorder which produces {\em well separated} Landau levels.   We find that a large  ratio 
 $\hbar\omega_c/\Gamma $, can produce large MIRO, and ZRS effects under radiation. It also is responsible for a large HIRO effect. Indeed these effects have been  observed in the same samples \cite{hiro1E}.

\item The conductivity at  $\omega=m\omega_c$ appears to be insensitive
to the radiation field (see Fig. \ref{fig:sigma-dom}).
Within Eq. (\ref{jx-asympt}), this is true only at temperatures $T/(\hbar \omega_c)=\nu^{max} >>1$,
otherwise the  Bessel functions sum rule,
\be
\lim_{\nu^{max}\to \infty}~\sum_{|\nu|\le \nu^{max} }  |J_\nu ( r \varepsilon_\omega )|^2 = 1,
\ee
is not saturated. At very low temperatures, only a few Bessel functions contribute, and we expect the conductivity at resonance frequencies to be suppressed  under radiation.

\item The characteristic field $E_\Gamma$  (\ref{EG1}), is  in the same ball park
as Ryzhii's estimate \cite{Ryzhii}. It  differs from the much larger estimate ($\ge \hbar\omega_c /(e R_c)$)
provided by Ref. \cite{Vavilov}.

\item The characteristic field  of  secondary  HIRO maxima (\ref{E-hiro}) depends on the disorder cut-off $q_s$. This is in contrast to
the semiclasssical picture \cite{hiro1A,VAG} of $2k_F$ scattering, or
touching cyclotron orbits of  radii   $R_c=v_F/\omega_c$.  This scenario is depicted in the inset of Fig. \ref{fig:K1-stat}. 
This argument  suggests enhanced scattering at an electric field  
\be
E_{2R_c}  ={ \hbar\omega_c \over  2e R_c} \ne E_{\omega_c}
\ee 
where our $E_{\omega_c}$  in (\ref{E-hiro}),  depends on $q_s$.
As  evidenced in  Figs. \ref{fig:WF}, and  \ref{fig:K1-stat}  there is no signature of  enhanced scattering at wavevectors  $ 2R_c/l_B^2$,  unless, by coincidence, the short range cut-off precisely
matches
$q_s = 2k_F$.  In summary, the  cyclotron radius is not
a noticeable  lengthscale in the  structure factor at high Landau levels. 

Experimentally \cite{hiro1C}, values of  oscillation field  have {\em not} been universal, and have varied in
the range $(1.63 , 2.18)  \times E_{2R_c}$.
\item An important issue, which we have ignored in this work,  is the relation between the dimensionless microwave field
 $\varepsilon_\omega$
and an external microwave radiation field and polarization. The resonance of positive circular polarization
at $\omega\!=\!\omega_c$ implies that close to the cyclotron resonance, one cannot ignore strong 
frequency and polarization dependence of the dielectric function.
There are  related open issues raised by  recent experiments, most notably the apparent independence of the MIRO on
microwave polarization\cite{smet1}, and the seeming inability of  2 photon absorption to explain 
a MIRO  about  $\omega_c/2$ \cite{smet2}.
We shall defer discussion of these problems to further investigations.
\end{enumerate}
 
{\em Acknowledgements.}
We thank Igor Aleiner, Herb Fertig, Shmuel Fishman, Leonid Glazman, Bert Halperin, Misha Raikh, Boris Shklovskii,
and Amir Yacoby  for informative discussions.
We acknowledge support from the Israel Science Foundation, the US Israel Binational Science Foundation, and the fund for Promotion of Research at Technion.
AA is grateful for the hospitality of Aspen Center for Physics.


\appendix
\section{The disorder matrix elements}
\label{App:disorder}
Using the definition (\ref{coordinates}),  
the  commutation between Landau operators and guiding center coordinates are readily obtained, for $\alpha,\beta=x,y$
 \bea
 \left[ \pi_\alpha, R_\beta \right] &=&0, \nonumber\\
\left[ \pi_x, \pi_y \right] &=& \left[ R_y, R_x \right] = i  .
\eea

The full disorder operator is expressed as
\be
\cV={1\over \cA} \sum_\bq V_\bq ~ e^{-i  l_B \bq\cdot\bPi \times{\hat\bz}} ~e^{-i l_B  \bq\cdot {\bR}} .
\ee

The matrix elements of the Fourier operators are given by
 \bea
 \langle k | e^{-i l_B  \bq\cdot {\bR}}  |k'\rangle&=&   e^{-i q_x k l_B} \delta_{k',k+q_yl_B}\nonumber\\
 \langle n| e^{-i  l_B \bq\cdot \bPi \times{\hat\bz}  }|m\rangle &=& e^{-(q l_B)^2\over 4} |{\bq l_B/ \sqrt{2}}|^{(m-n)}
 e^{i(m-n) \phi }  \nonumber\\
&&\times\sqrt{m ! \over n !}  L^{m-n}_{n} ((q l_B)^2/2)  \nonumber\\
&\equiv& D_{nm}(\bq) ,
\label{Dnm}
\eea
where $\phi=\arctan(q_y/q_x)$. Eq. (\ref{Dnm})  is used to split  the long and short wavelength terms in $\cV$
in Eqs. (\ref{Vknnp}) and (\ref{current-final}).

The asymptotic expression for the associated Laguerre polynomials\cite{GR} valid at large $n<m$, with 
$\tilde{q}=q l_B < \sqrt{2n}\pi$, yields
\be
\tilde{q} |D_{nm}(\tilde{q}^2/2)|^2 \sim {2\over \sqrt{2n}\pi} \cos^2(\sqrt{2n} \tq -\phi_{n-m}) ,
\label{asympt}
\ee
where $\phi= \pi/4 + \pi (m-n)/2$. The fast oscillations of the $\cos^2$ can be replaced
by its constant average of $\half$, which simplifies the $q$ integration over smooth functions considerably.

 \section{The Explicit Floquet Operator}
 \label{App:Floquet}
The full disordered evolution  operator  (\ref{UF}) is decomposed into
\be
U_F(t,t_0) = W(t)\exp\left( -{i\over \hbar} {\bar \cH}_0( t -t_0) \right)W^\dagger(t_0) ,
\ee
where ${\bar\cH}_0$ is the clean Landau level Hamiltonian,  and the vector-potential-dependent Floquet operator $W$ is defined as
 \be 
W(t) = e^{-i \bxi(t) \cdot\bPi } e^{- i \int^t {{L}(t')}dt'} .
\label{Wdef}
\ee
Here we shall determine the fields  $\bxi(t),L(t)$.
 $W$ acts on the time derivative,  and translates the Landau  level coordinates  $\bPi$ as follows
\begin{eqnarray}
&&W^\dag \left(   i\partial_t \right)   W=\nonumber\\
&& ~~\left( i\partial_t \right)+ \bPi\cdot{ \dot \bxi} 
 +\half \partial_t(\xi_x\xi_y) +L(t)   \nonumber \\
&&W^\dag \bPi  W  =\bPi+\bxi\times{\hat\bz}  \nonumber \\
&&W^\dag \br  W  = \br+ l_B \bxi    ,
\label{Floq-trans}
\eea
which transforms the   evolution equation as
\bea
&&W^\dag \left({i }\partial_t -{\omega_c \over 2} |\bPi+\ba|^2  \right) W \nonumber\\
&&~~=  {i}\partial_t -  {\omega_c \over 2}\left| \bPi+\bxi\times{\hat\bz} +\ba-{\dot\bxi}/\omega_c \right|^2\nonumber\\
&&  + { |{\dot\bxi}|^2 \over 2\omega_c} + \half(\xi_x\xi_y)+L(t) \nonumber\\
&& ~~=  {i\hbar \partial_t}-{\bar\cH}_d   .\nonumber\\
\label{Floq-def}
\eea

The fields $\bxi$ are given  by solving the linear equations
\bea
{\dot \bxi} - \omega_c \bxi\times{\hat\bz} &=&  \omega_c \ba(t) ,\nonumber\\
L(t)&=& \half(\xi_x\xi_y) .
\label{Floq-eqs}
\eea
For convenience, we fix the origin of time $t=0$, such that  $\bxi(t=0)=\bxi_0$.

Using the complex notation  ${\tilde\xi}=\xi_x+i\xi_y$, we write
\be
{\dot \txi} + i \omega_c \txi  = \omega_c {\tilde a}(t) .
\ee
The dimensionless gauge field (written in circular polarized components) is,
\bea
{\tilde a}&=& a_x+ia_y \nonumber\\
&=&  {e l_B\over\hbar  } \left(  {i E_\omega^+\over \omega}  e^{-i\omega t}  - {i E_\omega^-\over \omega}  e^{i\omega t} 
+ {\tilde E}_{dc}t  \right)  .
\eea
Separating the solution into  $\txi(t)=\txi^+ + \txi^- + \txi_0$, one obtains
\bea
\txi^{\pm} &=& \pm  { e l_B \omega_c \over \hbar \omega} {E_\omega^\pm \over  \omega_c \mp\omega}    e^{\mp i\omega t} ,\nonumber\\
\txi_0 &=& -i {e l_B{\tilde E}_{dc}\over\hbar  }  t - {e l_B{\tilde E}_{dc}\over\hbar\omega_c  } .
\label{txi}\eea
The zero frequency component of $\bxi$ in cartesian coordinates is
\be
\bxi_0(t) = t{e l_B \over\hbar  } {\bE}_{dc}\times\hat{\bz} - {e l_B\over\hbar\omega_c  }\bE_{dc} .
\label{bxi0}
\ee
In vector notation, we thus derived  the explicit form of the Floquet fields $\bxi(t)$
as given  in the main text by Eq.(\ref{etaxi}).


\begin{references}

\bibitem{Mani} R.G. Mani \etal., Nature {\bf 420}, 646 (2002);
\bibitem{Zudov1}M. A. Zudov, R. R. Du, J. A. Simmons, and J. R. Reno, Phys. Rev. B {\bf  64}, 201311(R) (2001).
\bibitem{Ye}P. D. Ye, L. W. Engel, D. C. Tsui, J. A. Simmons, J. R. Wendt, G.A. Vawter, and J. L. Reno, Appl. Phys. Lett. {\bf 79}, 2193 (2001).
\bibitem{Zudov2} M.A. Zudov, R.R. Du, L. N. Pfeiffer, K.W. West, Phys. Rev. Lett. {\bf 90}  046807 (2003).
\bibitem{Studenkin} S.A. Studenikin \etal, Phys. Rev. B {\bf 71} 245313 (2005).

\bibitem{hiro1A} C. L. Yang, J. Zhang, R. R. Du, J. A. Simmons, and J. L. Reno, Phys. Rev. Lett. {\bf 89}, 076801 (2002).
\bibitem{hiro1B} A. A. Bykov, J. Q. Zhang, S. Vitkalov, A. K. Kalagin, and A. K. Bakarov, Phys. Rev. B {\bf 72}, 245307 (2005).
\bibitem{hiro1C} W. Zhang, H.-S. Chiang, M.A. Zudov, L.N. Pfeiffer, K.W. West, cond-mat/0608727,  Phys. Rev. B {\bf 75}, 041304(R) (2007).

\bibitem{hiro1D} J-q. Zhang, S. Vitkalov, A. A. Bykov, A. K. Kalagin, A. K. Bakarov,  cond-mat/0607741.

\bibitem{hiro1E} W. Zhang, M. A. Zudov, L. N. Pfeiffer  and K.W. West,
Phys. Rev. Lett. {\bf  98}, 106804 (2007).

\bibitem{Andreev} A. V. Andreev, I. L. Aleiner, and A. J. Millis, Phys. Rev. Lett. {\bf 91},
056803 (2003).

\bibitem{A1} A. Auerbach, I.  Finkler, B. I. Halperin, and A.  Yacoby,  Phys. Rev. Lett. {\bf 94}, 196801 (2005).
\bibitem{A2} I. Finkler,  B.I. Halperin , A. Auerbach and A. Yacoby,  Jour. Stat. Phys. {\bf 125}, 1097 (2006), (cond-mat/0510722).



\bibitem{Ryzhii} V. I. Ryzhii, Fiz. Tverd. Tela {\bf 11}, 2577 (1969) [Sov. Phys.-Solid State {\bf 11}, 2078 (1970)];
V. Ryzhii and V. Vyurkov, Phys. Rev. B {\bf 68}, 165406 (2003);  V. I. Ryzhii,  cond-mat/0411370.
 
 \bibitem{Durst}  A.C. Durst, S. Sachdev, N. Read, S.M. Girvin,  Phys. Rev. Lett. {\bf 91},   086803 (2003).

\bibitem{Lei} X. L. Lei and S. Y. Liu,  Phys. Rev. Lett. {\bf 91}, 226805
(2003). 

\bibitem{Torres} M. Torres and A. Kunold, Phys. Rev. B {\bf 71},
115313 (2005);  J. Phys.: Cond. Matter {\bf 10}, 4029 (2006).

\bibitem{Vavilov} M. G. Vavilov and I. L. Aleiner,
Phys. Rev. B {\bf 69}, 035303 (2004).


\bibitem{Dmitriev} I. A. Dmitriev, M. G. Vavilov, I. L. Aleiner, A. D. Mirlin, and D. G. Polyakov,
Phys. Rev. B {\bf 71}, 115316 (2005).
 
\bibitem{VAG} M. G. Vavilov,  I. L. Aleiner,  and L. I. Glazman, Phys. Rev. B {\bf 76}, 115331 (2007).

\bibitem{AU} T. Ando and Y. Uemura, Jour. Phys. Soc. Japan {\bf 36}, 959 (1974).

\bibitem{comm:SCBA} The SCBA result for $\sigma^{xx}\propto (\tau_s/ \tau_{tr})$ \cite{RS} demonstrates that
perturbation theory about a clean system is singular in the long wavelength disorder. 

\bibitem{RS} M.E. Raikh and T.V. Shahbazyan, Phys. Rev. B {\bf 47}, 1522 (1993).

\bibitem{Dietel} Our  approach was inspired by the study of
MIRO effect in the presence of  a large periodic potential:  J. Dietel, L. I. Glazman, F. W. Hekking, and
F. von Oppen, Phys. Rev. B {\bf 71}, 045329 (2005).

\bibitem{Efros} A.L. Efros, F.G. Pikus, G.G. Samsonidze, Phys. Rev. B {\bf 41}, 8295 (1990).
\bibitem{Kohn}  W. Kohn, Phys. Rev. {\bf 123}, 1242 (1961).

\bibitem{Mehta} M. L. Mehta, ``Random Matrices'', Academic press (1991).


\bibitem{RMT} C. W. J. Beenakker,  Rev. Mod. Phys. {\bf 69}, 731  (1997).

\bibitem{comm-dark} (\ref{dark})   is smaller by a factor of $\sqrt{\pi}/4$ than the dark conductivity
of the  SCBA \cite{AU}, which is  probably due to the different form of density of states tails.

\bibitem{Fogler} M. M. Fogler, A. Yu. Dobin, and B. I. Shklovskii,
Phys. Rev. B {\bf 57}, 4614  (1998).

\bibitem{Balents} J. Alicea, L. Balents, M.P.A. Fisher, A. Paramekanti, L. Radzihovsky,
Phys. Rev. B {\bf 71}, 235322 (2005).

\bibitem{smet1} J. H. Smet, B. Gorshunov, C. Jiang, L. Pfeiffer, K. West, V. Umanksy, M. Dressel, R. Meisels, F. Kuchar, K. von Klitzing,  cond-mat/0505183 

\bibitem{smet2} S. I. Dorozhkin, J. H. Smet, K. von Klitzing, L. N. Pfeiffer, K. W. West, 
cond-mat/0608633


\bibitem{GR} I.S. Gradshteyn and I.M. Ryzhik, "Table of Integrals, series and [products",
(Academic press, 1980); page 1039.





\end{references}
\end{document}